\documentclass[12pt,preprint]{aastex}
\usepackage{color}
\shorttitle{Chromospheric emission of planet candidate systems -- a way to identify false positives}
\shortauthors{Karoff et al.}

\begin{document}

\title{Chromospheric emission of planet candidate host stars -- a way to identify false positives}

\author{Christoffer Karoff}
\affil{Department of Geoscience, Aarhus University, H{\o}egh-Guldbergs Gade 2, 8000, Aarhus C, Denmark}
\affil{Stellar Astrophysics Centre, Department of Physics and Astronomy, Aarhus University, Ny Munkegade 120, 8000, Aarhus C, Denmark}
\email{karoff@phys.au.dk}

\author{Simon Albrecht}
\affil{Stellar Astrophysics Centre, Department of Physics and Astronomy, Aarhus University, Ny Munkegade 120, 8000, Aarhus C, Denmark}

\author{Alfio Bonanno}
\affil{INAF - Osservatorio Astrofisico di Catania, via S.Sofia 78, 95123, Catania, Italy}

\author{Mads Faurschou Knudsen}
\affil{Department of Geoscience, Aarhus University, H{\o}egh-Guldbergs Gade 2, 8000, Aarhus C, Denmark}

\begin{abstract}
It has been hypothesized that the presence of closely orbiting giant planets is associated with enhanced chromospheric emission of their host stars. The main cause for such a relation would likely be enhanced dynamo action induced by the planet. We present measurements of chromospheric emission in 234 planet candidate systems from the {\it Kepler} mission. This ensemble includes 37 systems with giant planet candidates, which show a clear emission enhancement. The enhancement, however, disappears when systems which are also identified as eclipsing binary candidates are removed from the ensemble. This suggests that a large fraction of the giant planet candidate systems with chromospheric emission stronger than the Sun are not giant planet system, but false positives. Such false-positive systems could be tidally interacting binaries with strong chromospheric emission. This hypotesis is supported by an analysis of 188 eclipsing binary candidates that show increasing chromospheric emission as function of decreasing orbital period.
\end{abstract}

\keywords{stars: activity -- stars: chromospheres  -- stars: solar-type -- binaries: eclipsing -- planets-star interactions}

\section{Introduction}
Although we cannot observe stellar magnetic fields directly, we can observe how they affect the atmospheres around stars. In the chromosphere, the injection of magnetic energy leads to a deviation from radiative equilibrium. The effect of this deviation from radiative equilibrium can be observed as emission in i.e.\ the Ca {\sc ii} H \& K spectral lines at 396.8 and 393.4~nm, respectively. The intensity of this emission scales with the amount of non-thermal heating in the chromosphere, making these lines a useful proxy for the strength of, and fractional area covered by, magnetic fields. This was first suggested by \citet{1913ApJ....38..292E} and has subsequently been used extensively to measure stellar activity since Olin Wilson started regular observations at the Mount Wilson Observatory \citep{1968ApJ...153..221W, 1978ApJ...226..379W, 1995ApJ...438..269B}. 

The discovery of the Jupiter-mass companion 51~Peg\,b in a close orbit around it's solar-type host star \citep{1995Natur.378..355M} 
 inspired a number of studies which suggested that hot Jupiters (HJ) could lead to enhanced magnetic activity 
\citep{2000ApJ...529.1031R, 2000ApJ...533L.151C, 2004ApJ...602L..53I, 2005ApJ...622.1075S, 2008A&A...487.1163L, 2009A&A...505..339L, 2012A&A...540A..82K}. 
In particular, \citet{2000ApJ...533L.151C} argued that the presence of HJs can enhance the magnetic activity of their host stars 
either through tidal interactions or through direct magnetic interaction. 
Both these effects are expected to increase chromospheric, transition region, and coronal activity, although the avaiable statistical evidence remains a matter of debate \citep[see e.g.][]{2010A&A...515A..98P, miller2015}.

The only mechanism that could decrease the apparent measured chromospheric activity of exoplanet host stars is the enhanced photometric stellar noise of these stars, which would decrease the detection likelihood of exoplanets as this leads to additional noise in transit and Radial Velocity (RV) data \citep{2002ApJ...575..493J,2003A&A...401..743C}. However, this effect is likely not significant for giant planets, as their transit and RV signals are larger than the variability caused by chromospheric activity. Given our current understanding of giant planets, especially giant planets in short period orbits, we would thus expect to see enhanced chromospheric activity in systems with HJs compared to an ensemble of solar-like stars without HJs. 

Another reason that can explain why systems with HJ candidates might show higher chromospheric emission, is misclassification. If the giant-planet systems are misclassified and the companions are not close-in giant planets, but low-mass stars, then the HJ candidates will show higher chromospheric emission, as stars in close-in binary star system can show enhanced chromospheric emission originating from tidal interactions \citep[see i.e.][]{1977A&A....57..383Z, 1991A&A...251..183S}.

Here, we analyse data from an ensemble of Kepler Objects of Interest (KOIs), some of which harbour planets and some which are suspected to harbour a stellar companion, to test if the enhanced emission is due to misclassification or to star-planet interactions.

\section{Observations}
\citet{karoff2016} used observations from the Large Sky Area Multi-Object Fibre Spectroscopic Telescope \citep[{\sc LAMOST}, also called the Guoshoujing Telescope at Xinglong observatory, China,][]{2012RAA....12.1243L} to measure the chromospheric emission in the Ca {\sc II} H \& K spectral lines at 396.8 and 393.4~nm, respectively, for 5648 main-sequence, solar-like stars with effective temperatures between 5100 and 6000~K. Here, we build on this study by extending the temperature range from the coolest main-sequence stars up to 6500 K, which means that the stellar ensemble is extended to include 13263 main-sequence stars. Although {\sc LAMOST} has a relative low resolution of only 1800 and the chromospheric emission is measured in spectra with a signal-to-noise ratio down to 10 in the blue part, the analysis by \citet{karoff2016} showed that the chromospheric emission measurements were reliable. In this study, we restrict ourselves to stars cooler than 6500 K for two reasons. Firstly, stars above the Kraft break \citep{1967ApJ...150..551K} hardly have any outer convection zones and are therefore not expected to have a magnetic dynamo like the Sun. Secondly, the H$\zeta$ line at 388.9~nm starts to be prominent in stars above the Kraft break -- leading to a dramatic increase in the S index as the V reference band used to measure the S index is located between 389.1 and 391.1~nm. In the cool end, the required signal-to-noise ratio means that almost all our stars are hotter than 5000 K.

For our analysis, we use information on the $Kepler$ Objects of Interests (KOI) available through the Exoplanet Archive \citep[][downloaded on 2016 May 23]{2013PASP..125..989A}. The KOI list includes 2290 confirmed planets and 2416 planetary candidates. We also downloaded the Kepler Eclipsing Binary Catalog \citep[][downloaded on 2016 January 7]{2016AJ....151...68K}. This catalog contains 2878 eclipsing binaries.

\section{Analysis}
Cross-correlating the list of 13263 main-sequence, solar-like stars with the KOI list results in an ensemble of 234 stars that have confirmed or candidate planets and well-determined chromospheric emission. We define a giant planet (GP) candidate as a planet with a radius larger than 6 times Earth's radius (R$_{\earth}$), finding that 37 KOIs fulfil this criterium. These 37 GP candidates include 18 HJ candidates with orbital periods shorter than 10 days (see Table 1). 

If we consider the confirmed planets, the cross-correlation results in 123 planets, with 10 being defined as GPs and 5 as HJs. 

The chromospheric emission is estimated by \citet{karoff2016} as both the S index \citep{1978PASP...90..267V}, and the chromospheric flux through the two quantities $R_{\rm HK}^+$ and $F_{\rm HK}^+$ \citep{2013A&A...549A.117M}. The reason for calculating the chromospheric flux is that the S index is known to be color dependent \citep{1984ApJ...279..763N} and that the S index is a purely empirical quantity, without a physical unit. The problem is, however, that additional information like B-V color or effective temperature is needed in order to convert the S index into chromospheric flux. For the analysis in \citet{karoff2016}, effective temperatures from the $Kepler$ Input Catalog (KIC) were used to calculate the chromospheric fluxes, which resulted in larger noise associated with the chromospheric flux estimates, than in the chromospheric emission measurements. For the present study we, as in \citet{karoff2016}, base the analysis on the S index, but use the chromospheric fluxes for consistency check. 

For systems with multiple planets we only look at the largest planet, assuming that the mass is the most important parameter for star-planet interactions. The semi-major axis could have been chosen too, but we find that small planets are general not associated with enhanced chromospheric emisssion.

\section{Results}
We compared relative distributions (Fig.~1 panel A) of the chromospheric emission for all the 13263 main-sequence, solar-like stars  (black, denoted the general ensemble), for all the 234 KOIs (red, denoted the planet ensemble) and for the 37 KOIs with GP candidates (purple). The distributions of all  main-sequence, solar-like stars and the 234 KOIs appears to be almost identical. This is also reflected by the fact that the mean S index of the two distributions, with $0.2084\pm0.0004$ and $0.2069\pm0.0028$ respectively, are almost identical (see Table~1). On the other hand, the distribution of the KOIs with GP candidates is significantly different from the distribution of all the main-sequence, solar-like stars. The GP distribution has a mean S index of $0.2277\pm0.0078$. A Kolmogorov-Smirnov test indicates that the S indices of the main-sequence stars and the S indices of the GP ensemble are drawn, at the 4$\sigma$ confidence level, from two different distributions (p value 0.991). 

We repeated the analysis described above, but only using KOIs which are not also listed in the Kepler Eclipsing Binary Catalog as eclipsing binary candidates. There appears to be no significant difference in the S index between the three distributions (general, planet and GP, see Fig.~1 panel B). This is also confirmed by a Kolmogorov-Smirnov test (Table~1). When we only consider HJ candidates, we observe the same picture, i.e. the apparent enhancement of chromospheric emission of KOIs with GP candidates disappears when the eclipsing binary candidates are removed from the ensemble. 

\paragraph{Chromospheric flux} 
In order to test this result, we performed the same analysis using the chromospheric flux, $R_{\rm HK}^+$. These calculations confirm the results obtained using the S index (see Table~2). KOIs with GP candidates have higher chromospheric flux than solar-like stars in general, but  this excess in chromospheric flux vanishes when eclipsing binary candidates are removed from the ensemble.

\paragraph{Anderson-Darling test} 
The Kolmogorov-Smirnov test evaluates the relation between the empirical distribution functions of the two ensembles being tested. Despite its many advantages, the Kolmogorov-Smirnov test may have some short-comings, when the mean value of the two ensembles are similar, but the empirical distribution functions differ. In that case, the Anderson-Darling test can provide a more reliable test, as it gives more weight to the tails of the empirical distribution functions \citep{anderson1952}. We have therefore used an Anderson-Darling test to validate the results of the Kolmogorov-Smirnov tests in Tables~1 \& 2. This exercise confirms that although there are small differences between the significance returned by the Kolmogorov-Smirnov and the Anderson-Darling tests, none of the conclusions depend on the choice of test.   

\paragraph{Bootstrap test} 
Given the moderate ensemble sizes, it is a possibility that the apparent enhancement of chromospheric emission of KOIs with GP candidates disappears when the eclipsing binary candidates are removed from the ensemble due to the decrease in sample size. In order to test this we performed a bootstrap test, where we randomly removed stars from the planet, GP and HJ ensembles. The number of stars that were removed is the same as number of stars in the three ensembles that were found to be on the eclipsing binary candidate list.  The test was performed 10000 times and returned mean p values of 0.766, 0.999 and 0.999 for the difference between the general ensemble and the planet, GP and HJ ensembles, respectively. We thus conclude that the reduction is sample size cannot account for the disappearance of the observed enhancement of chromospheric emission when the eclipsing binary candidates are removed from the ensemble.

\paragraph{Effective temperature of the different ensembles} 
Although we both calculate the S index and $R_{\rm HK}^+$ for all stars, there is a possibility that differences in the distributions of the effective temperature of any of the ensembles could lead to a difference in any of the two indicies. We therefore calculated the mean effective temperature for the general, planet, GP and HJ ensembles and found they were not different within the associated uncertainties. A Kolmogorov-Smirnov test also revealed that the effective temperatures of the planet, GP and HJ ensembles were not significantly different from the general ensemble. The ensemble of the eclipsing binary candidates were, however, shifted to higher temperatures compared to the general ensemble (p value of 0.989). This is not unexpected, as the binary fraction is known to increase with increasing effective temperatures \citep[see e.g.][]{2014ApJ...788L..37G}. It is, however, seen in Fig. ~2, which shows the distribution of chromospheric emission for all the 13263 main-sequence stars, but here separated into stars cooler and hotter than the Sun, that the hotter stars show slightly stronger chromospheric activity than the cooler stars. The reason for this is likely the H$\zeta$ line, as dynamo theory predicts that stars cooler than the Sun should be magnetically more active than stars hotter than the Sun. The same result appears when the $R_{\rm HK}^+$ is considered-- indicating that this discrepancy is not caused by improper calibration of the S index.

\section{Discussions}
Our analysis of chromospheric emission in 13263 main-sequence, solar-like stars, including 234 confirmed and candidate planet host stars shows that a) KOIs with GP candidates have higher chromospheric emission than main-sequence, solar-like stars in general, but b) when eclipsing binary candidates are removed from the list of KOIs with GP candidates, the apparent enhancement of chromospheric emission of KOIs with GP candidates vanish. This can be explained in a scenario where the main part of the GP candidates orbiting main-sequence, solar-like stars with strong chromospheric emission are not GPs, but tidally interacting binaries. Support for this scenario is found in Fig.~2, where we also compare the distributions of chromospheric emission of 188 eclipsing binary candidates. These are separated into candidates with orbital periode below and above 10 days. Clear enhancement of the chromospheric emission is particularly visible for eclipsing binary candidates in close-in orbits. This can be attributed to tidal interactions between the two components \citep{1977A&A....57..383Z, 1991A&A...251..183S}.  Due to the larger masses, tidal interactions will be stronger in star-star systems than in star-planet systems. Fig.~3 shows the chromospheric emission for the 188 eclipsing binary candidates and the 37 KOIs with GP candidates as a function of orbital period. Here, the eclipsing binary candidates show clear decreasing chromospheric emission with increasing orbital periods, especially for orbital period below 10 days. This relation is much weaker for the KOIs with GP candidates. This result is confirmed when looking at $R_{\rm HK}^+$.

\paragraph{Confirmed planets} 
The proposed scenario, where the GP candidates orbiting main-sequence, solar-like stars with strong chromospheric emission are in fact not giant planets, but tidally interacting binaries, also suggests that the apparent enhancement of chromospheric emission for KOIs with GP candidates should disappear when only looking at confirmed GP planets. This is in fact also what we see in Table~1. None of the confirmed planet, GP or HJ ensembles are significantly different (p value $>$ 0.95) from the general ensemble. There are however a few cases where the p value is above 0.90, especially for the planet ensemble. In part, this could be due to the fact that it is easier to confirm a planet round an inactive star than around an active one. This is also supported by the fact that the mean values of the S index show very little difference between the planet ensemble and the general ensemble. Similarly, as indicated in Fig.~3, eclipsing binary candidates show very little enhancement of chromospheric emission for orbital periods longer than 10 days. We would also expect the apparent enhancement of chromospheric emission for KOIs with GP candidates should diminish when only considering GPs that are not HJs. This is also what we observe, with a mean value of the S index for the GP with orbital period longer than 10 days of $0.2215\pm0.0113$, when not excluding eclipsing binaries and $0.2134\pm0.0123$ when excluding binaries. Also, the probability for the two ensembles being significantly different from the general ensemble is only 0.740 and 0.355, respectively.  

\paragraph{False-positive rate} 
If the scenario that the main part of the GP candidates orbiting main-sequence, solar-like stars with strong chromospheric emission are not giant planets, but tidally interacting binaries is true then we can use our estimate to calculate the false-positive rate. This is done by computing how many, randomly chosen, eclipsing binary candidates we would need to include in an ensemble of 37 and 18 stars selected from the general ensemble in order to have an ensemble that is as significantly different from the general ensemble as the GP and HJ ensembles, respectively. This analysis, which was repeated 10000 times, revealed that we would need 16 eclipsing binary candidates in the GP ensemble and 11 in the HJ ensemble. Our analysis thus indicates a false-positive rate of $56.8\%$ for GP and $63.6\%$ for HJ candidates found by $Kepler$. 

Our estimates of the false-positive rate are not proper measurements based on e.g. a six-year RV campaign as done by \citet{2016A&A...587A..64S}, who finds a false-positive rate of $54.6\pm6.5\%$. The agreement between our estimates and the false-positive rate measure by \citet{2016A&A...587A..64S} does, however, support the scenario proposed in this study.

\citet{2016A&A...587A..64S} distinquish between eclipsing binaries and contaminating eclipsing binaries, where $37.2\pm5.4\%$ of the GP candidates are in fact eclipsing binaries and $15.1 \pm 3.4\%$ are contaminating eclipsing binaries. Although our analysis cannot discriminate between eclipsing binaries and contaminating eclipsing binaries, it is likely that a significant fraction $(>10\%)$ of the planetary candidates, as well as the eclipsing binary candidates, are in fact background objects (objects that fall on the same line-of-sight as the brightest star). This means that the true chromospheric activity distribution of the eclipsing binary ensemble would be shifted to even higher values and our estimated false-positive rates would be lower.

Although our study does not provide evidence for enhancement of chromospheric emission of GP host stars, we cannot rule out the possibility that GPs can lead to enhanced emission of their host star in some systems. Based on this study, we can only conclude that GPs in general do not lead to a chromospheric enhancement of their hosts stars larger than a $\sim$0.01 difference in the S index  \citep[see i.e.][]{2014A&A...565L...1P}. 
 
\section{Conclusions}
We analyse chromospheric emission measured in 13263 main-sequence stars including 234 confirmed and candidate planet host stars. Our analysis revealed that 37 KOIs with GP candidates show a clear enhancement of chromospheric emission. The enhancement, however, disappears when KOIs that are also listed as eclipsing binary candidates are removed from the sample. This suggests that the enhancement of chromospheric emission is due to the presence of tidally interacting binaries with strong chromospheric emission rather than star-planet interactions caused by the presence of a giant planet. We thus use the chromospheric emission of 188 eclipsing binaries to show how chromospheric emission of eclipsing binaries increase with decreasing orbital period.

If the enhancement of chromospheric emission of KOIs with GP candidates is indeed caused by stellar companions, it suggests a false positive rate of a $\sim56.8\%$ for $Kepler$ GP candidates and $\sim63.6\%$ for HJ candidates.

\acknowledgments
We would like to thank the anonymous referee for valuable comments that help improve our manuscript substantially. Guoshoujing Telescope (the Large Sky Area Multi-Object Fiber Spectroscopic Telescope, LAMOST) is a National Major Scientific Project built by the Chinese Academy of Sciences. Funding for the project has been provided by the National Development and Reform Commission. LAMOST is operated and managed by the National Astronomical Observatories, Chinese Academy of Sciences. This research has made use of the NASA Exoplanet Archive, which is operated by the California Institute of Technology, under contract with the National Aeronautics and Space Administration under the Exoplanet Exploration Program. CK thanks section A30 at Aarhus University Hospital where part of this manuscript was written. Funding for the Stellar Astrophysics Centre is provided by the Danish National Research Foundation (Grant agreement No.: DNRF106). The project has been supported by the Villum Foundation and the Danish Council for Independent Research, through a DFF Sapere Aude Starting grant No. 4181-00487B. 

\appendix

\begin{figure}
\epsscale{1.0}
\plotone{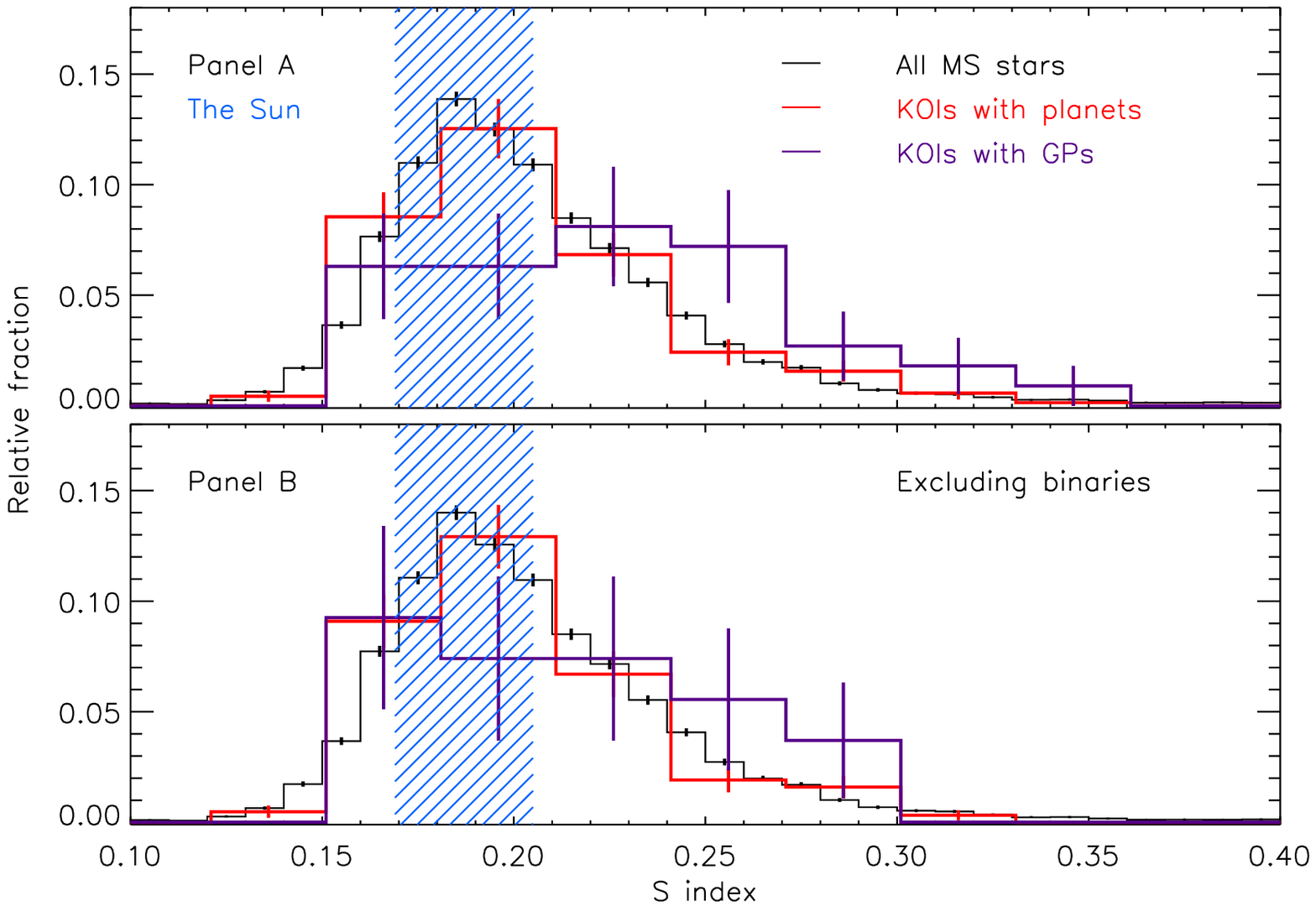}
\caption{Comparison of measured distributions of chromospheric emission. Panel A compare the distributions of chromospheric emission in main-sequence, solar-like stars (black line) to the distributions of these stars, which have confirmed or candidate planets around them (red line) and the distribution of these stars, which have confirmed or candidate GP around them (purple line). The blue shaded region marks the range of the S index of the Sun between solar cycle minima and maxima \citep{1996AJ....111..439H}. The error bars are based on the number of stars in each bin, assuming Poisson statistics. The uncertainties on the measured S indecies are smaller than 0.03 \citep{karoff2016}. It is seen that KOIs with GP candidates have higher chromospheric emission than main-sequence, solar-like stars in general. Panel B shows the same as the panel A,  but here the candidate eclipsing binaries have been removed from the emsembles. It is seen that when eclipsing binaries are removed from the list of KOIs, the apparent enhancement of chromospheric emission of KOIs with GP candidate seen in panel A vanish.}
\end{figure}

\begin{figure}
\epsscale{0.6}
\plotone{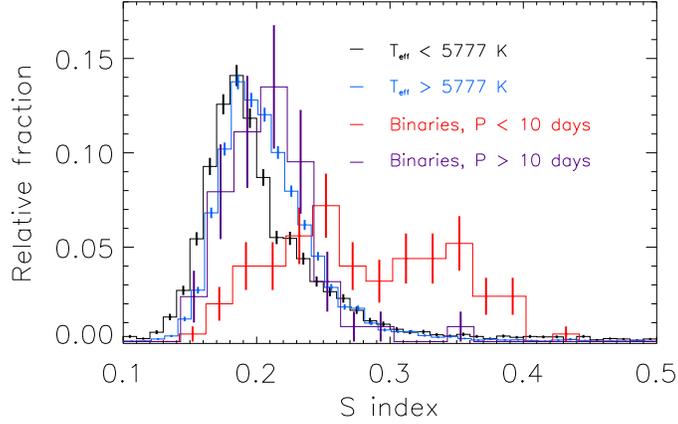}
\caption{Comparison of measured distributions of chromospheric emission for hot and cool main-sequence stars as well as candidate eclipsing binaries. It is seen that hot main-sequence stars tend show slightly stronger chromospheric emission than cool main-sequence stars. This is likely due to the H$\zeta$ line. It is also seen that close-in eclipsing binary candidates show strong chromospheric emission. This is likely due to tidal interactions.}
\end{figure}

\begin{figure}
\epsscale{0.6}
\plotone{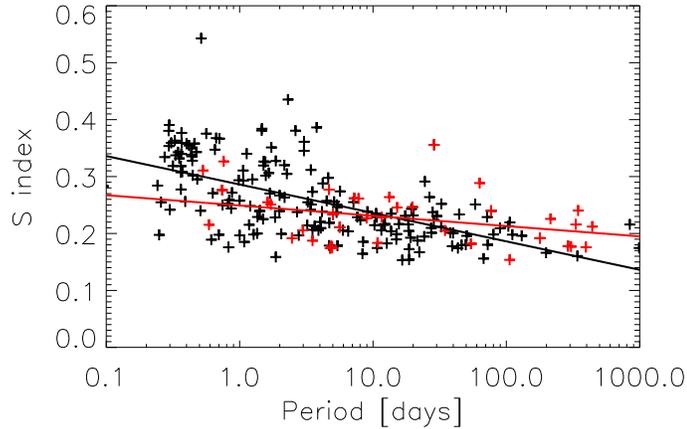}
\caption{Relation between chromospheric emission and orbital period for eclipsing binary candidates (black) and KOIs with GP candidates (red). For the eclipsing binary candidates a generally increasing trend is seen for the chromospheric emission with decreasing orbital period. Especially for eclipsing binary candidates with orbital periods below 10 days. This is likely caused by tidal interactions.  The black line shows a log-linear fit to the eclipsing binary candidates $S=0.286\pm0.005-0.050\pm0.005{\rm log}P$. This trend is much weaker for the KOIs with GP candidates $S=0.250\pm0.012-0.018\pm0.008{\rm log}P$ (red line).}
\end{figure}

\begin{table}
\caption{Statistical significance of the difference in S index. K-S Stat. stands for Kolmogorov-Smirnov statistic; K-S Sig. for significance of the Kolmogorov-Smirnov test; A-D Sig. for significance of the of the Anderson-Darling test; P for planet; GP for giant planet and HJ for hot Jupiter.}
\centering
\begin{tabular}{lrlccc}
\hline \hline
\vspace{-0.5cm}\\
 & Number of stars &  Mean ${\rm S}$ index & K-S Stat. & K-S Sig. & A-D Sig. \\
\hline
\multicolumn{3}{l}{Including eclipsing binary candidates} & & & \\
\hline
All stars                         & 13263 & 0.2084(4)  &  & & \\
P confirmed		    &     123 & 0.2047(42) &   0.086 & 0.689 & 0.856\\
P candidates 		    &     234 & 0.2069(28) &   0.041 & 0.174 & 0.365\\
GP confirmed		    &       10 & 0.2051(95) & 0.185 & 0.153 & 0.129\\
GP candidates 	   	    &       37 & 0.2277(78) &   0.264 & 0.991 & 0.997\\
HJ confirmed                &         5 & 0.1985(179) & 0.389 & 0.654 & 0.404\\
HJ candidates               &       18 & 0.2342(112) &   0.326 & 0.966 & 0.993\\
\hline
\multicolumn{3}{l}{Excluding eclipsing binary candidates.} & & &\\
\hline
All stars                         & 12075 & 0.2077(4)  &  &  & \\
P candidates 		    &     209 & 0.2041(29) &   0.064 & 0.638 & 0.803\\
GP candidates 	            &       18 & 0.2129(94) &   0.200 & 0.575 & 0.375\\
HJ candidates 	             &       7 & 0.2138(171) &   0.199 & 0.085 & 0.210\\
\hline
\end{tabular}
\label{tab1}
\end{table}

\begin{table}
\caption{Statistical significance of the difference in $R^+_{\rm HK}$ The abbreviations are the same as in Table 1.}
\centering
\begin{tabular}{lrlccc}
\hline \hline
 &  Number of stars & Mean $R^+_{\rm HK}$ & K-S Stat. & K-S Sig. & A-D Sig.\\
\hline
\multicolumn{3}{l}{Including eclipsing binary candidates} & & &\\
\hline
All stars                         & 13239 & -4.621(2)  &  &\\
P confirmed                   &    123  & -4.624(16) & 0.113 & 0.917 & 0.944 \\
P candidates 		    &     234 & -4.623(11) &   0.052 & 0.437 & 0.715\\
GP confirmed                &       10  & -4.605 (40) & 0.189 & 0.172 & 0.322\\
GP candidates 	             &      37 & -4.534(30) &   0.293 & 0.997 & 0.996\\
HJ confirmed                &         5 &        -4.661(55) & 0.385 & 0.640 & 0.407\\
HJ candidates 	            &       18 & -4.513(39) &   0.311 & 0.952 & 0.990\\
\hline
\multicolumn{3}{l}{Excluding eclipsing binary candidates} & & & \\
\hline
All stars                         &  13051 & -4.624(2)  &  &\\
P candidates 		    &     209 & -4.634(11) &   0.073 & 0.791 & 0.926\\
GP candidates 	    	   &     18 & -4.585(41) &   0.208 & 0.621 & 0.431\\
HJ candidates 	           &       7 & -4.589(64) &   0.233 & 0.213 & 0.159\\
\hline
\end{tabular}
\label{tab2}
\end{table}

\end{document}